\documentclass[aps,noshowpacs,floatfix,superscriptaddress,preprintnumbers,showpacs,showkeys,nofootinbib]{revtex4}
\usepackage{graphicx}
\usepackage{amsmath,amssymb}
\usepackage{subfigure,epsfig}
\usepackage{slashed,braket}
\usepackage{bbold,bm}

\newcommand{\be}{\begin{equation}}
\newcommand{\ee}{\end{equation}}
\newcommand{\ba}{\begin{eqnarray}}
\newcommand{\ea}{\end{eqnarray}}

\newcommand{\eq}{\begin{eqnarray}}
\newcommand{\en}{\end{eqnarray}}
\begin{document}

\title{Exploring hyperon structure with electromagnetic transverse densities
}

\author{Jose Manuel Alarc\'on}
\affiliation{Theory Center, Jefferson Lab, Newport News, VA 23606, USA}
\author{Astrid N.~Hiller Blin}
\affiliation{Institut f\"ur Kernphysik \& PRISMA Cluster of Excellence, Johannes Gutenberg Universit\"at, D-55099 Mainz, Germany}
\author{Manuel J.~Vicente Vacas}
\affiliation{Instituto de F\'{\i}sica Corpuscular, 
Universidad de Valencia--CSIC,\\
Institutos de Investigaci\'on, Ap. Correos 22085, E-46071 Valencia, Spain}
\author{Christian Weiss}
\affiliation{Theory Center, Jefferson Lab, Newport News, VA 23606, USA}

\today

\begin{abstract}
We explore the structure of the spin-1/2 flavor-octet baryons (hyperons) through
their electromagnetic transverse densities. The transverse densities describe the 
distribution of charge and magnetization at fixed light-front time and enable a 
spatial representation of the baryons as relativistic systems. 
At peripheral distances $b \sim 1/M_\pi$ the transverse densities are
computed using a new method that combines chiral effective field theory ($\chi$EFT) 
and dispersion analysis. The peripheral isovector densities arise from two-pion exchange,
which includes the $\rho$ resonance through elastic unitarity. The isoscalar densities 
are estimated from vector meson exchange ($\omega, \phi)$. We find that the ``pion cloud''
in the charged $\Sigma$ hyperons is comparable to the nucleon, while in the $\Xi$ it
is suppressed. The $\Lambda$--$\Sigma^0$ transition density is pure isovector and
represents a clear manifestiation of peripheral two-pion dynamics.
\end{abstract}

\maketitle

\section{Introduction}
\label{Sec:Introduction}
Understanding the structure of strange baryons is an important goal of hadronic physics. 
The $SU(3)$ octet baryons (hyperons) are stable under strong interactions and possess 
an electromagnetic and weak structure similar to that of the nucleon, in terms of 
vector and axial current matrix elements, which can be measured in radiative 
transitions and weak decays \cite{Lach:1995we,Cabibbo:2003cu}. 
It is thus possible to characterize the hyperons
by charge and current densities and compare these to those in the nucleon.
Interesting questions are whether the hyperons are more ``compact'' or more ``extended''
than the nucleon, and how they couple to the chiral degrees of freedom responsible
for long-range structure (``pion cloud''). The answers to these questions have
implications also for the understanding of hyperon-hyperon interactions and the 
role of strangeness in strong interaction dynamics at low energies; see 
Ref.~\cite{Dover:1985ba} for a review.

For relativistic systems such as hadrons the electromagnetic structure can be
expressed in terms of transverse densities. They are defined as the 2-dimensional
Fourier transforms of the hadron form factors and describe the spatial distribution
of charge and current in the system at fixed light-front time $x^+ = x^0 + x^3 =$ 
const \cite{Soper:1976jc,Burkardt:2000za,Burkardt:2002hr,Miller:2007uy}.
As such they are boost-invariant and provide an objective representation of
the hadron as an extended system. They are closely related to the partonic description 
of hadron structure in QCD and correspond to a projection of the generalized 
parton distributions (GPDs). Transverse densities have been used extensively in
studies of nucleon structure; see Ref.~\cite{Miller:2010nz} for a review of results.
They can equally well be used to explore hyperon structure and answer the 
above questions.

At peripheral distances $b\sim 1/M_\pi$ the transverse densities can be computed 
model-independently using a new method that combines chiral effective field theory
($\chi$EFT) and dispersion analysis \cite{Alarcon:2017asr}. The densities are represented 
as dispersive integrals over the imaginary parts of the baryon form factors on the cut 
in the timelike region, $\textrm{Im} \, F^B(t)$ at $t > t_{\rm thr}$. The spectral 
functions on the two-pion cut ($t_{\rm thr} = 4 M_\pi^2$) are constructed using
the elastic unitarity condition and the $N/D$ method, with dynamical input 
from $\chi$EFT and the timelike pion form factor measured in $e^+e^-$ annihilation
experiments \cite{Alarcon:2017asr,Granados:2017cib}. 
The method effectively includes the $\rho$ meson resonance in the 
$\pi\pi$ channel, which plays an essential role in electromagnetic structure. 
It permits calculation of the isovector peripheral densities down to distances 
$b \gtrsim 1\, \textrm{fm}$ with controled accuracy. In this article we review 
the results of the method for the hyperon densities and their impact on the
understanding of peripheral hyperon structure. Further applications of the method 
are described in Refs.~\cite{Alarcon:2017ivh,Alarcon:2017lhg}.
\section{Formalism}
\label{Sec:Transverse_densities}
The matrix element of the electromagnetic current between spin-1/2 baryon states
with 4-momenta $p$ and $p'$
is described by two form factors, $F_1^B(t)$ and $F_2^B(t)$ (Dirac and Pauli form factors).
They are functions of the invariant momentum transfer $t = \Delta^2 = (p' - p)^2$ and can be
measured and interpreted without specifying a particular form of relativistic dynamics
or reference frame. In the light-front form of relativistic dynamics one follows the 
evolution of strong interactions in light-front time $x^+ \equiv x^0 + x^3$
\cite{Dirac:1949cp,Leutwyler:1977vy,Brodsky:1997de}. In this
context it is natural to consider the form factors in a frame where the 4-momentum
transfer has only transverse components $\bm{\Delta}_T = (\Delta^x, \Delta^y), 
|\bm{\Delta}_T|^2 = -t$, and to represent them as Fourier transforms of 
two-dimensional spatial densities
\begin{align}
\label{Eq:rho_def}
F_i^B (t = -|\bm{\Delta}_T|^2) \;\; = \;\; \int d^2 b \; 
e^{i \bm{\Delta}_T \cdot \bm{b}} \; \rho_i^B (b) \hspace{2em} (i = 1, 2),
\end{align}
where $\bm{b} \equiv (b^x, b^y)$ is a transverse coordinate variable and $b \equiv |\bm{b}|$.
The functions $\rho_1^B(b)$ and $\rho_2^B(b)$ describe the transverse spatial distribution of 
charge and magnetization in the baryon at fixed $x^+ = 0$ and are invariant under boosts
in the $z$-direction. Their interpretation as spatial 
densities and other properties have been discussed extensively in the 
literature \cite{Burkardt:2000za,Burkardt:2002hr,Miller:2010nz,Granados:2013moa}. 
Alternative to the magnetization density $\rho_2^B(b)$ one also considers the function
\begin{align}
\widetilde\rho_2^B (b) & \equiv \frac{\partial}{\partial b} 
\left[ \frac{\rho_2^B (b)}{2 m_B} \right] ,
\label{rho_2_tilde_def}
\end{align}
which has a simple partonic interpretation. Together, $\rho_1^B(b)$ and $\widetilde\rho_2^B(b)$
contain the full information about the current matrix element and provide a concise spatial 
representation of the baryons' electromagentic structure.

The baryon form factors are analytic functions of $t$ and have a dispersive representation
of the form
\begin{align}
F_i^B (t) \;\; = \;\; 
\int_{t_{\rm thr}}^\infty \frac{dt'}{t' - t - i0} 
\; \frac{\textrm{Im}\, F_i^B (t')}{\pi} 
\hspace{2em} (i=1,2) ,
\label{Eq:ff-spectral-rep}
\end{align}
in which they are expressed as integrals over the imaginary parts (spectral functions) on 
the cut at $t > t_{\rm thr}$. The spectral functions arise from processes in which the 
current produces a hadronic state that couples to the baryon-antibaryon system,
current $\rightarrow$ hadronic state $\rightarrow$ $B\bar B$.
Using this representation in Eq.~(\ref{Eq:rho_def}), one
obtains a dispersive representation of the densities \cite{Strikman:2010pu,Miller:2011du} 
\begin{align}
& \rho_1^B (b) = \phantom{-} \int_{t_{\rm thr}}^{\infty} \! dt\ \frac{K_0(\sqrt{t}b)}{ 2\pi}
\frac{\text{Im}F_1^B(t)}{\pi} ,
\label{Eq:rho1-spectral-rep} \\
& \widetilde{\rho}_2^B (b) = - \int_{t_{\rm thr}}^{\infty} \! dt\ \frac{\sqrt{t}
K_1(\sqrt{t}b)}{ 4\pi m_B} \frac{\text{Im} F_2^B(t)}{\pi}.
\label{Eq:rho2tilde-spectral-rep}
\end{align}
Here $K_n \; (n = 0, 1)$ are the modified Bessel functions, which decay exponentially
at large arguments, $K_n (\sqrt{t} b) \sim (\sqrt{t} b)^{-1/2} \, e^{-\sqrt{t} b}$
for $\sqrt{t} b \gg 1$. The integrals for the densities therefore converge exponentially 
at large $t$. The distance $b$ determines at what values of $t$ the spectral function is
effectively sampled in the integral (``exponential filter''). In particular, the densities
at large distances are governed by the lowest-mass hadronic states in the spectral function.
In the isovector densities this is the two-pion state (threshold $t_{\rm thr} = 4\, M_\pi^2$), 
which includes the $\rho$ resonance at $t \sim 30\, M_\pi^2$; in the isoscalar 
densities these are effectively the $\omega$ and $\phi$ resonances in the $3\pi$ and 
$K\bar K$ channels. The densities thus enable a parametric definition of ``peripheral''
baryon structure and relate it to the spectral decomposition of the form factors.

\begin{figure*}
\begin{center}
\epsfig{file=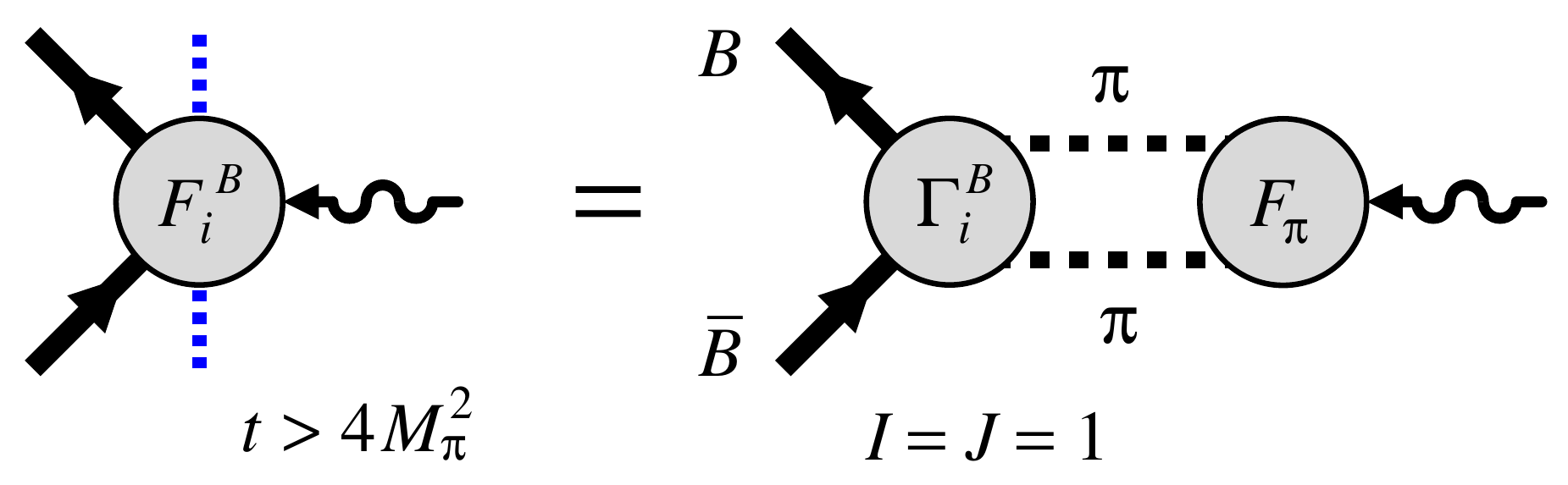,width=.6\textwidth,angle=0} 
\end{center}
\caption[]{Unitarity relation for the isovector spectral function on the two-pion cut.}
\label{fig:unitarity}
\end{figure*}
The isovector spectral functions on the two-pion cut can be computed in a new approach 
based on elastic unitarity, the $N/D$ method, and dynamical input 
from $\chi$EFT and timelike pion form factor measurements \cite{Alarcon:2017asr}. 
The elastic unitarity condition in the two-pion 
channel allows one to express the spectral functions 
as \cite{Frazer:1960zza,Frazer:1960zzb,Hohler:1974ht}
\begin{equation}
\label{Eq:Disp_rep_ImFV}
\textrm{Im} F_i^B(t) = \frac{k_{\rm cm}^3}{\sqrt{t}} \; \Gamma_i^B(t) \; F^*_{\pi}(t) 
\hspace{2em} (i=1,2),
\end{equation}
where $k_{\rm cm}=\sqrt{t/4 - M_\pi^2}$ is the center-of-mass momentum of the $\pi\pi$ system
in the $t$-channel, $\Gamma_i^B(t)$ is the $I=J=1$  $\pi \pi \rightarrow B\bar{B}$ partial wave 
amplitude, and $F_{\pi}(t)$ is the pion timelike form factor (see Fig.~\ref{fig:unitarity}). 
The complex amplitudes $\Gamma_i^B(t)$ and $F_{\pi}(t)$ have the same phase on the two-pion cut;
the phase arises from $\pi\pi$ rescattering in the $t$-channel, which affects both amplitudes
in the same way (Watson theorem) \cite{Watson:1954uc}. The unitarity condition 
Eq.~(\ref{Eq:Disp_rep_ImFV}) can thus be written in manifestly real form as
\begin{equation}
\label{Eq:Disp_rep_ImFV_ratio}
\textrm{Im} F_i^B(t) = \frac{k_{\rm cm}^3}{\sqrt{t}} \; \frac{\Gamma_i^B(t)}{F_{\pi}(t) } \; 
|F_\pi(t)|^2 \hspace{2em} (i=1,2).
\end{equation}
The ratio $\Gamma^B_i (t) / F_\pi(t)$ is real and free of $\pi\pi$ rescattering effects.
This function can be computed in $\chi$EFT with relativistic baryons with controled accuracy. 
The factor $|F_\pi(t)|^2$ contains the $\pi\pi$ rescattering effects and the $\rho$ resonance,
and is taken as the empirical form factor measured in $e^+e^-$ annihilation experiments.
The approach allows us to construct the two-pion spectral functions of baryons in the region
$4 M_\pi^2 < t \lesssim 1\, \textrm{GeV}^2$, which includes the $\rho$ meson resonance.
Further aspects of the method are discussed 
in Refs.~\cite{Alarcon:2017asr,Alarcon:2017lhg,Alarcon:2017ivh}. 

\begin{figure*}
\begin{center}
\epsfig{file=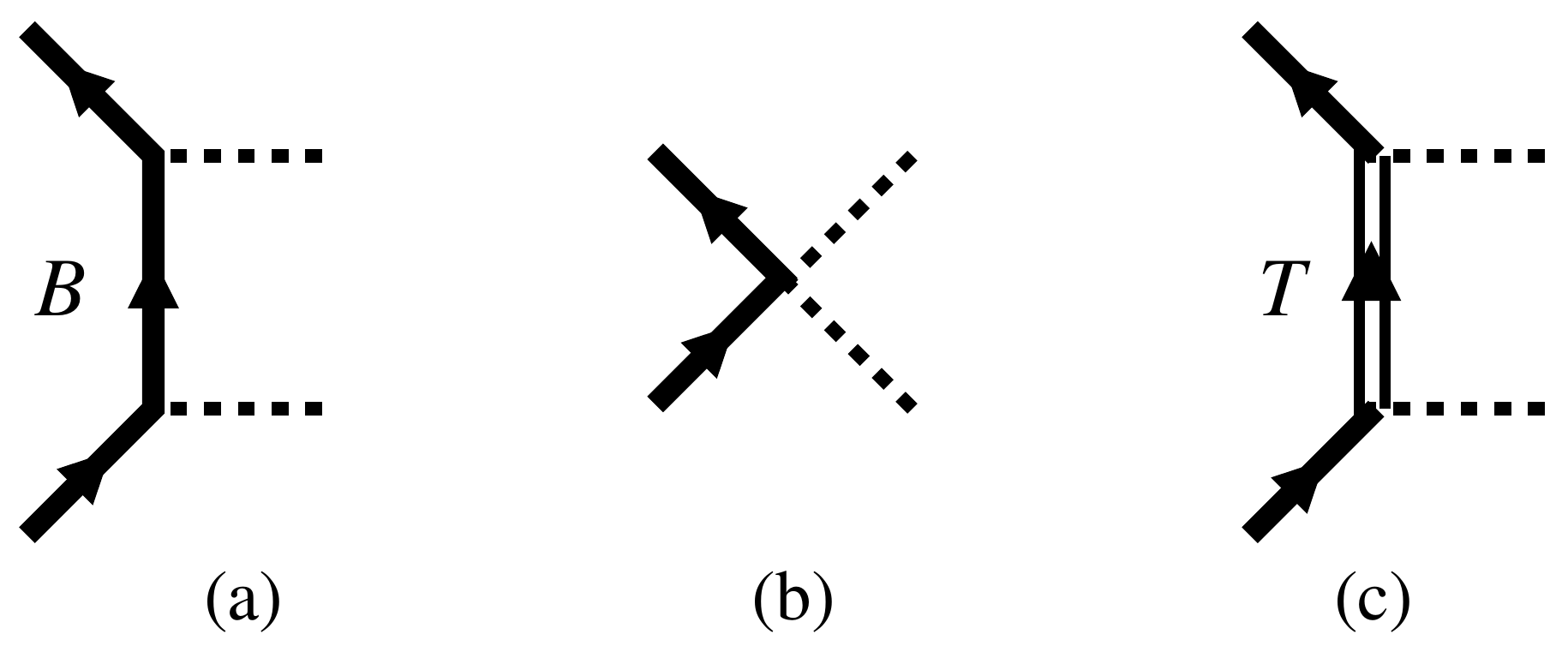,width=.55\textwidth,angle=0} 
\end{center}
\caption[]{LO $\chi$EFT diagrams contributing to the $\pi\pi \rightarrow N\bar N$
partial-wave amplitudes $\Gamma_i^B$ in the $I = J = 1$ channel. (a)~Born term
with intermediate octet baryon $B$. (b)~Weinberg-Tomozawa contact term. (c)~Born term
with intermediate decuplet baryon $T$.
\label{fig:eft}}
\end{figure*}

The isovector spectral functions of the $SU(3)$ octet baryons have been calculated with the
above method, using relativistic $\chi$EFT with spin-1/2 octet and spin-3/2 decuplet baryons 
in LO accuracy \cite{Alarcon:2017asr}. The chiral processes contributing to
partial-wave amplitudes $\Gamma_i^B(t)$ at this accuracy are shown in Fig.~\ref{fig:eft}.
[Calculations in the $SU(2)$ sector
have meanwhile been extended to NLO and partial N2LO accuracy and show good convergence 
in higher orders \cite{Alarcon:2017lhg}.] The isoscalar spectral functions have been modeled
by vector meson exchange ($\omega, \phi$), with couplings constrained by $SU(3)$ symmetry
and dispersive fits to the nucleon form factor data \cite{Alarcon:2017asr}.\footnote{The 
contribution of $K\bar K$ states to the hyperon isovector and isoscalar spectral functions
in $\chi$EFT was computed in Ref.~\cite{Blin:2017hez} without rescattering effects.
The contributions of these states to the peripheral densities considered here turns out 
to be negligible.}
With these spectral functions we have evaluated the peripheral transverse densities of the hyperons
through the dispersive representation Eqs.~(\ref{Eq:rho1-spectral-rep}) 
and (\ref{Eq:rho2tilde-spectral-rep}).
\section{Results and discussion}
\label{Sec:Transverse_densities_octet}
\begin{figure*}
\begin{center}
\epsfig{file=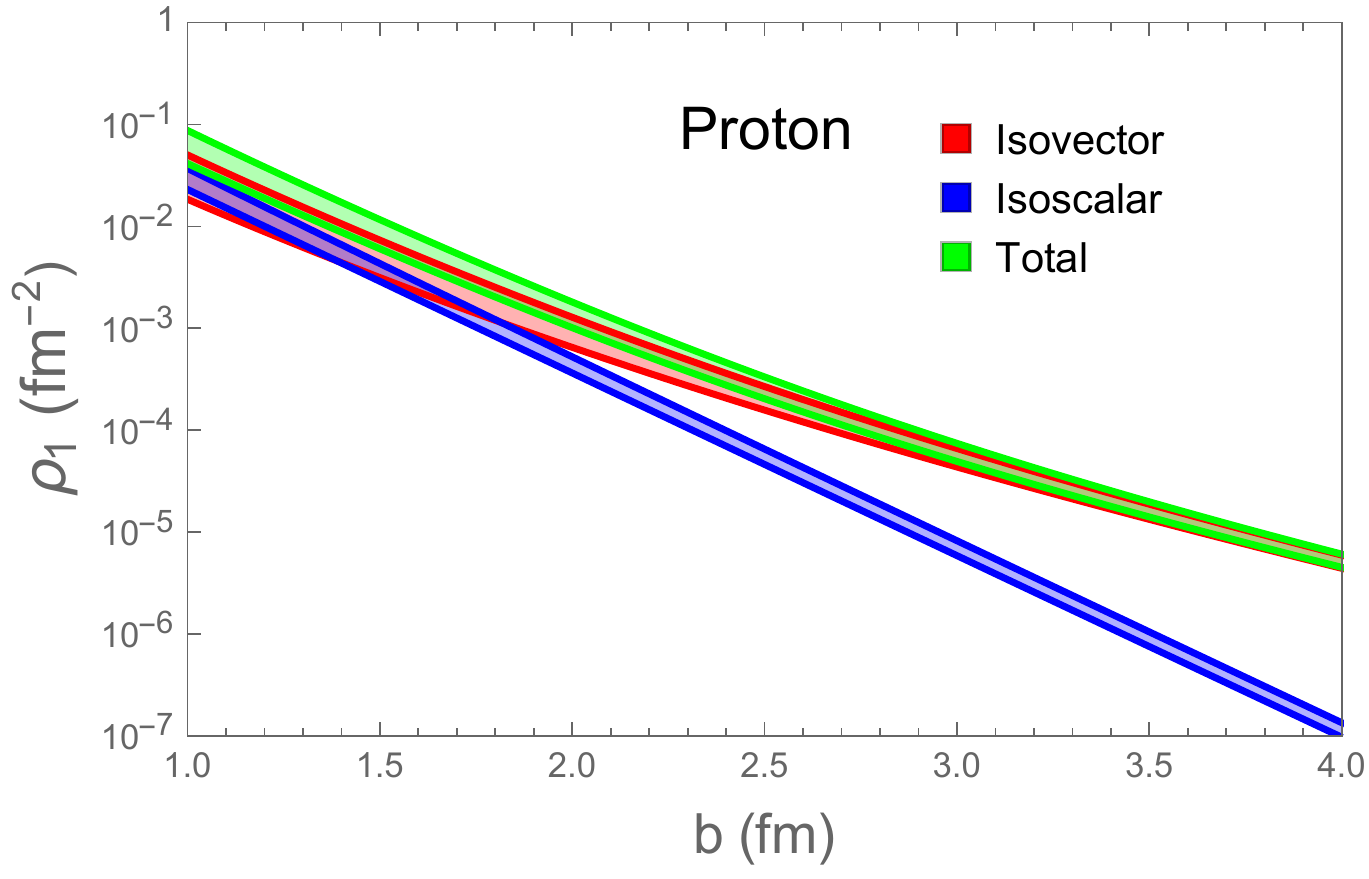,width=.45\textwidth,angle=0}
\epsfig{file=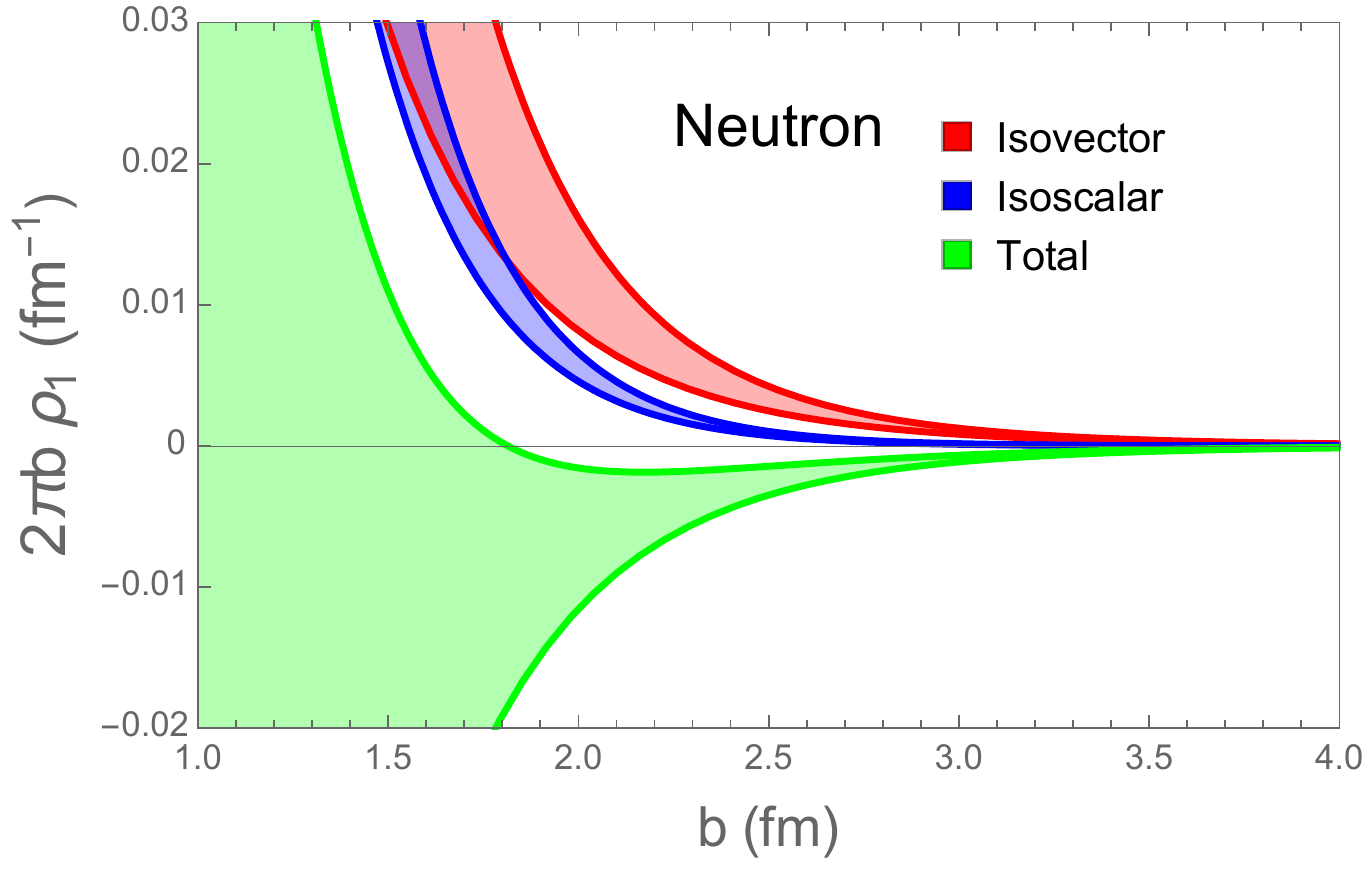,width=.45\textwidth,angle=0}\\
\epsfig{file=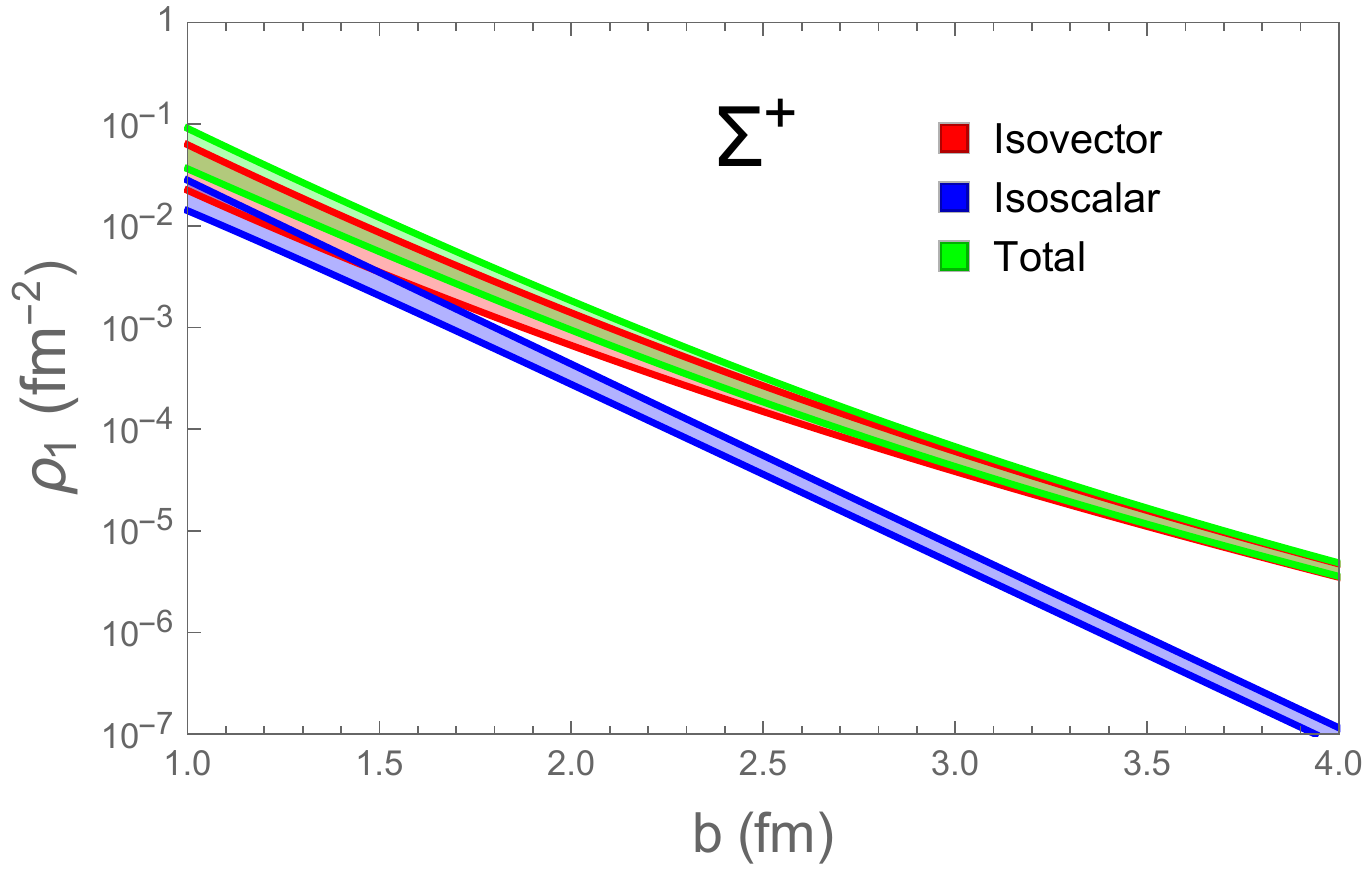,width=.45\textwidth,angle=0}
\epsfig{file=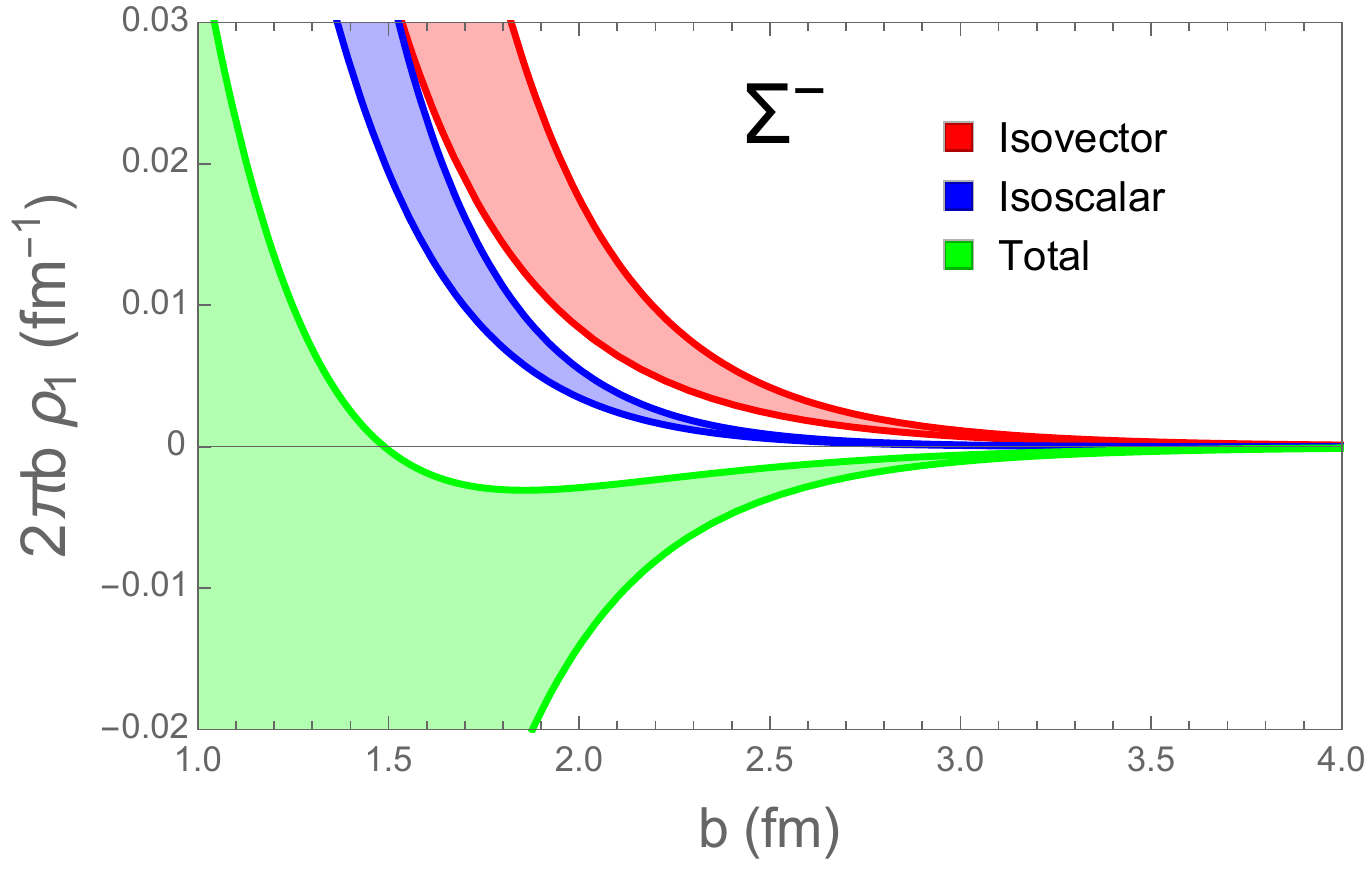,width=.45\textwidth,angle=0}\\
\epsfig{file=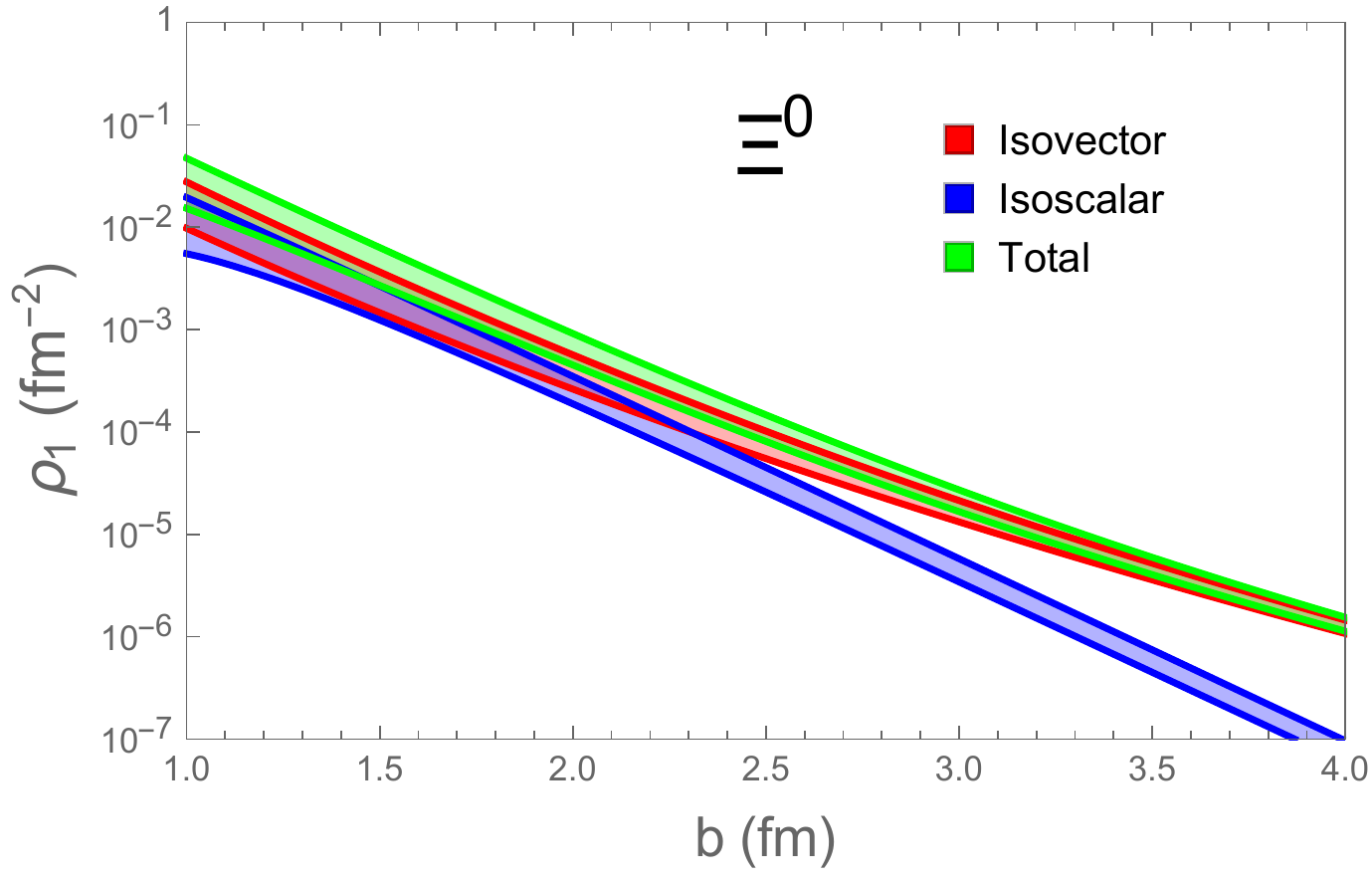,width=.45\textwidth,angle=0}
\epsfig{file=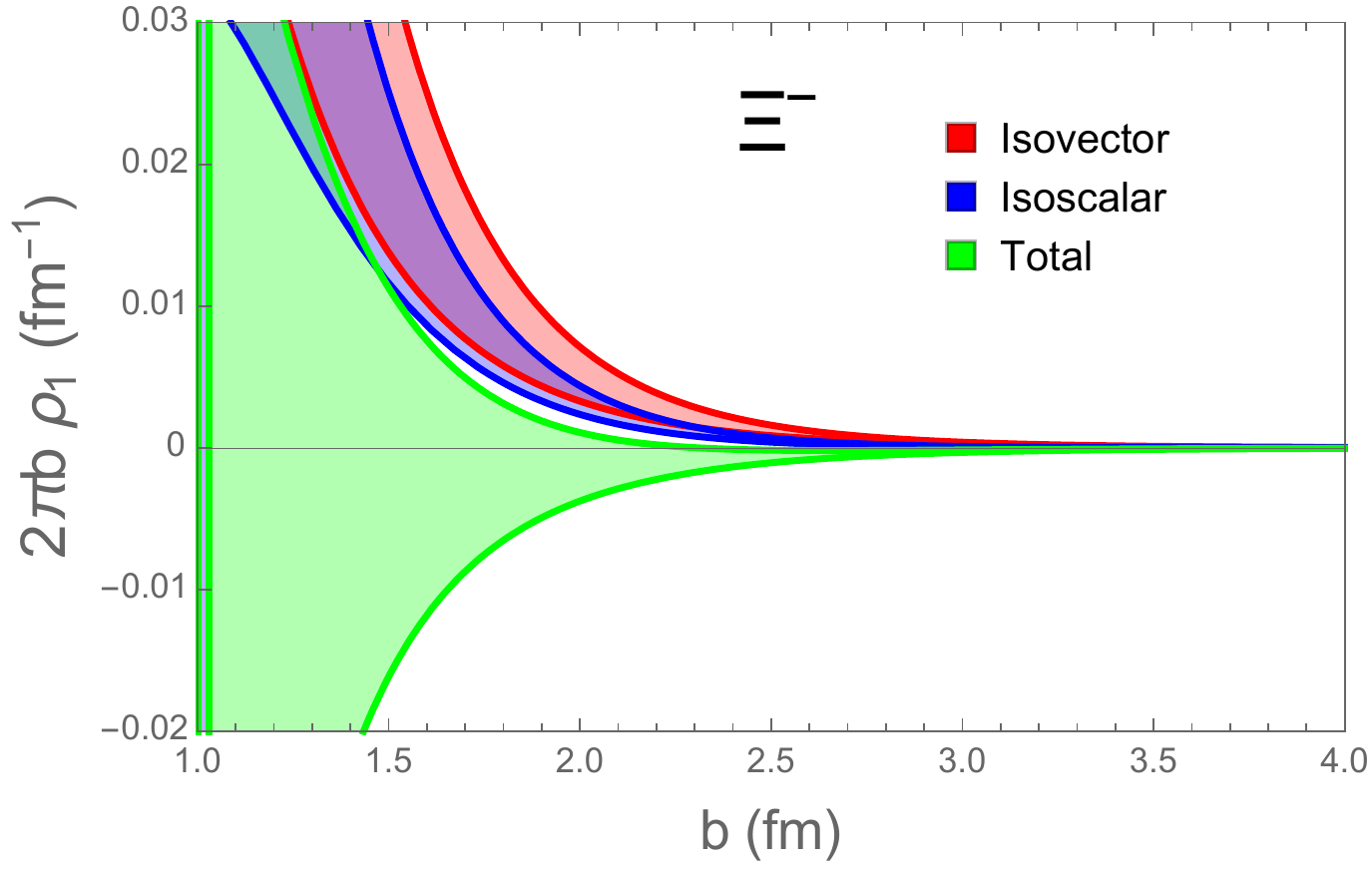,width=.45\textwidth,angle=0} \\
\epsfig{file=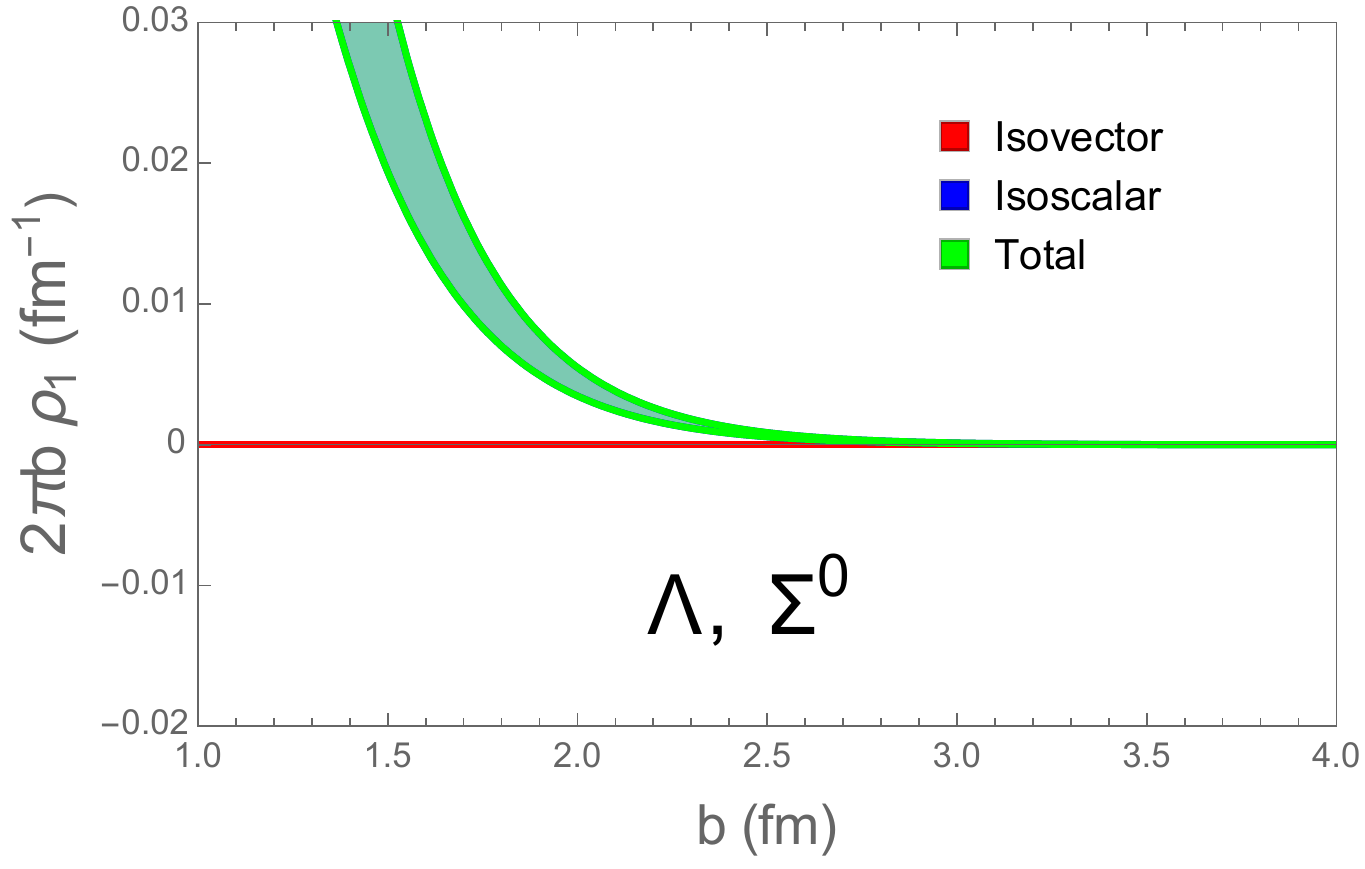,width=.45\textwidth,angle=0}
\epsfig{file=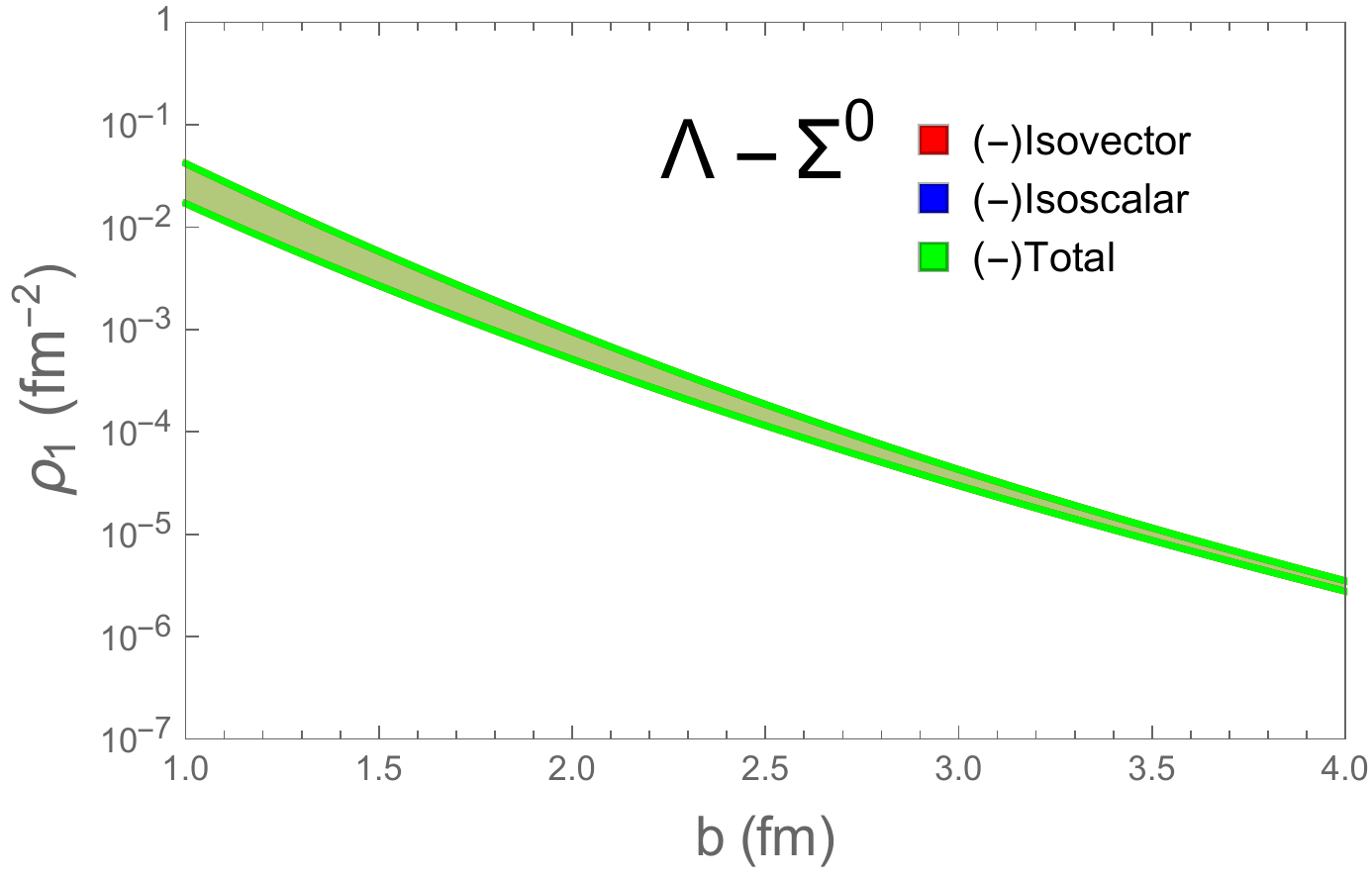,width=.45\textwidth,angle=0}
\caption[]{Peripheral transverse charge densities of the octet baryons \cite{Alarcon:2017asr}. 
Red: Isovector component calculated with the spectral functions obtained from
Eq.~(\ref{Eq:Disp_rep_ImFV_ratio}), $\chi$EFT, and the empirical pion form factor.
Blue: Isoscalar component estimated from vector meson poles. 
Green: Total density (sum or difference of isoscalar
and isovector components). For the densities with fixed sign we
plot $\rho_1(b)$ on a logarithmic scale (the signs are indicated in the legends of the
plots); for those with changing sign we plot the
radial densities $2\pi b \rho_1(b)$ on a linear scale.}
\label{Fig:rho1-Octet}
\end{center}
\end{figure*} 
%
%
\begin{figure*}
\begin{center}
\epsfig{file=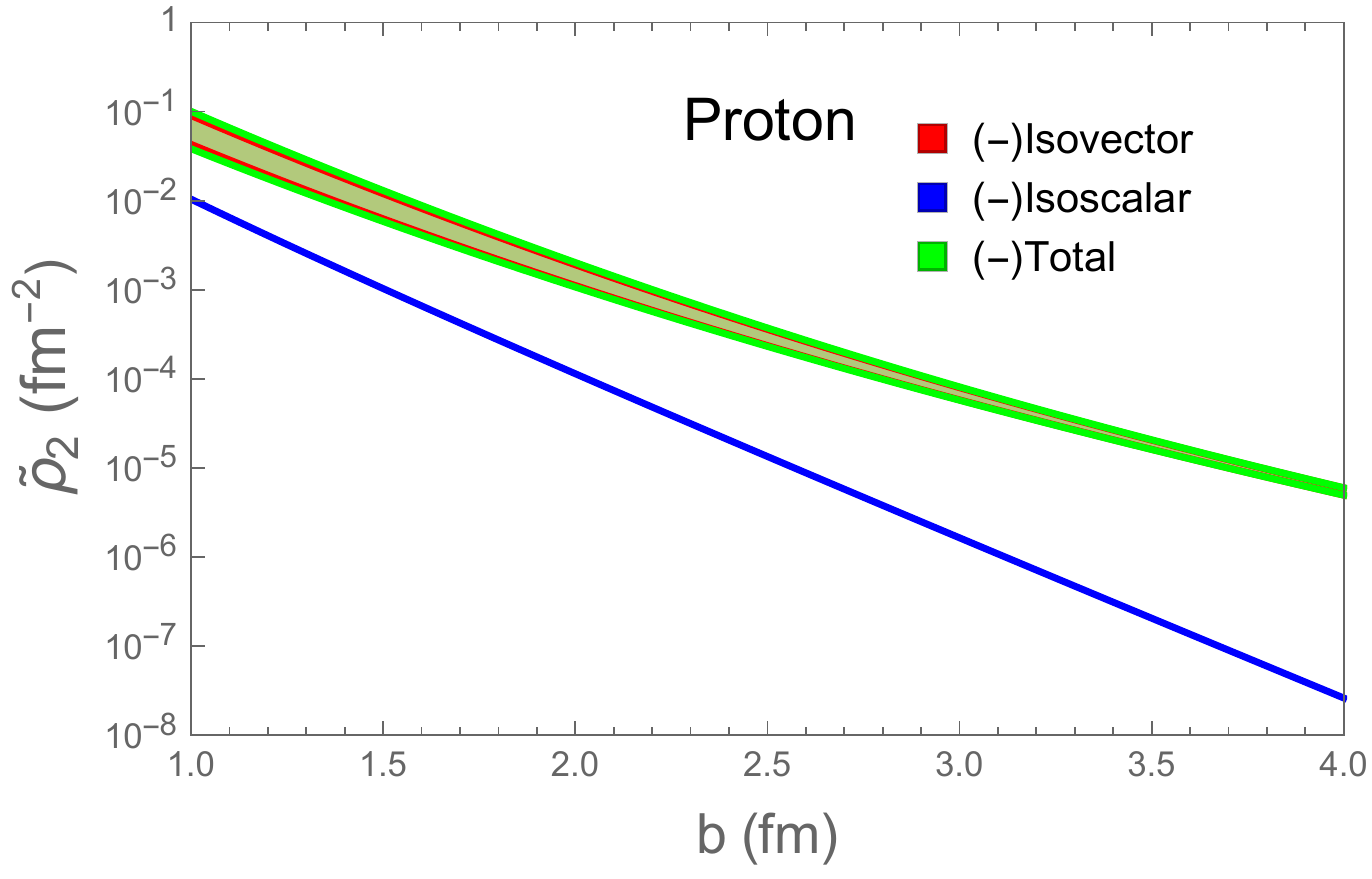,width=.45\textwidth,angle=0}
\epsfig{file=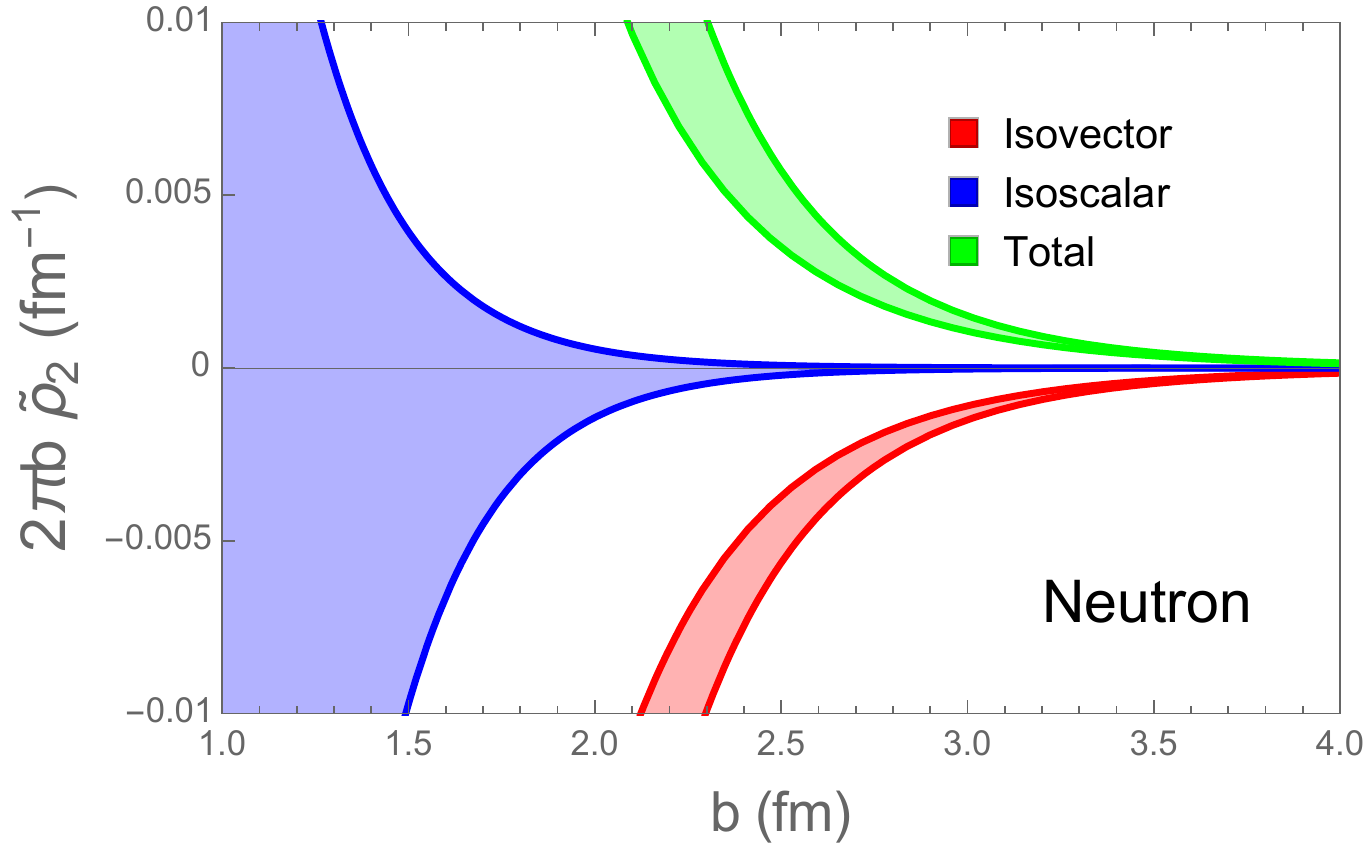,width=.45\textwidth,angle=0}\\
\epsfig{file=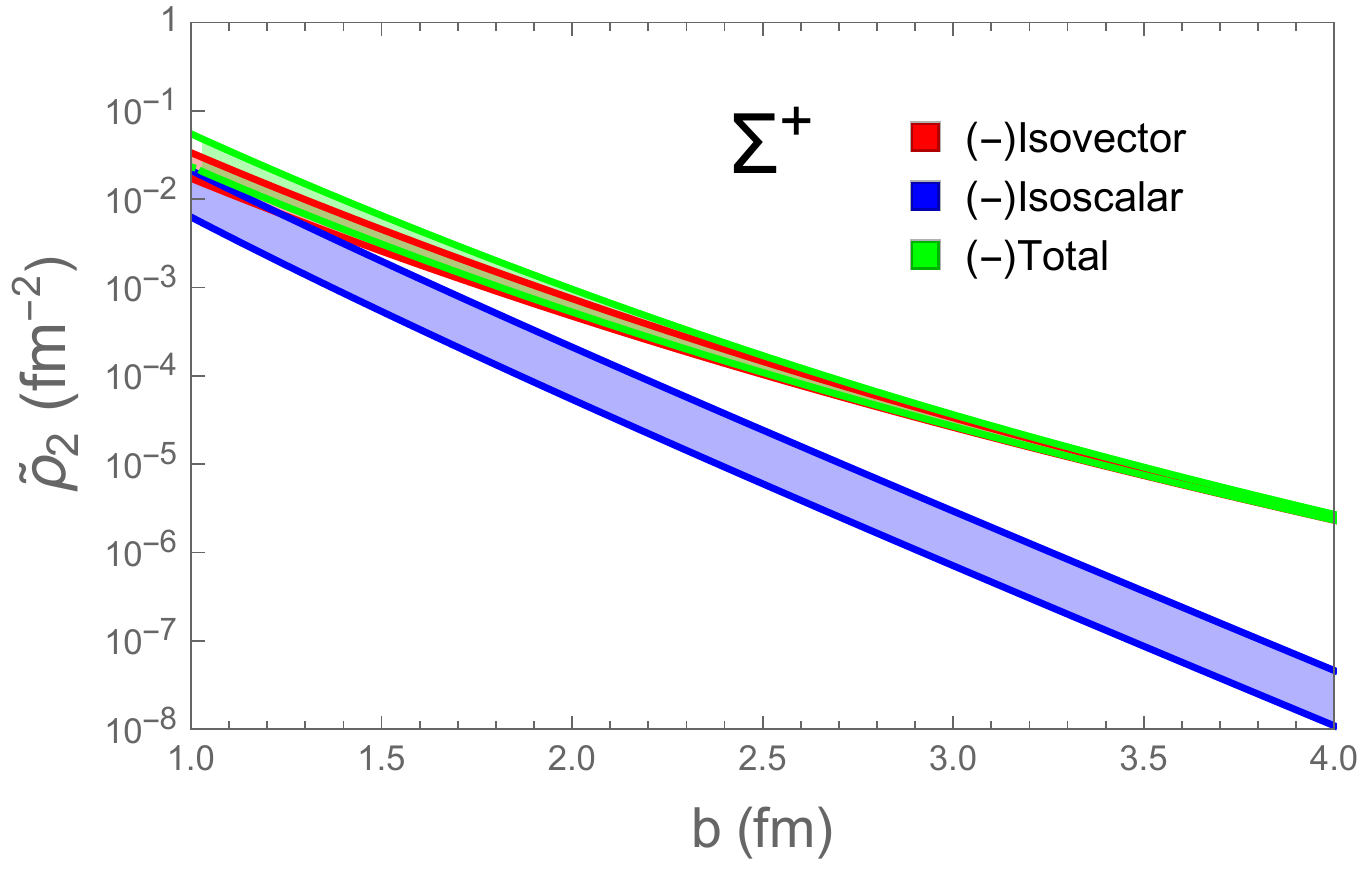,width=.45\textwidth,angle=0}
\epsfig{file=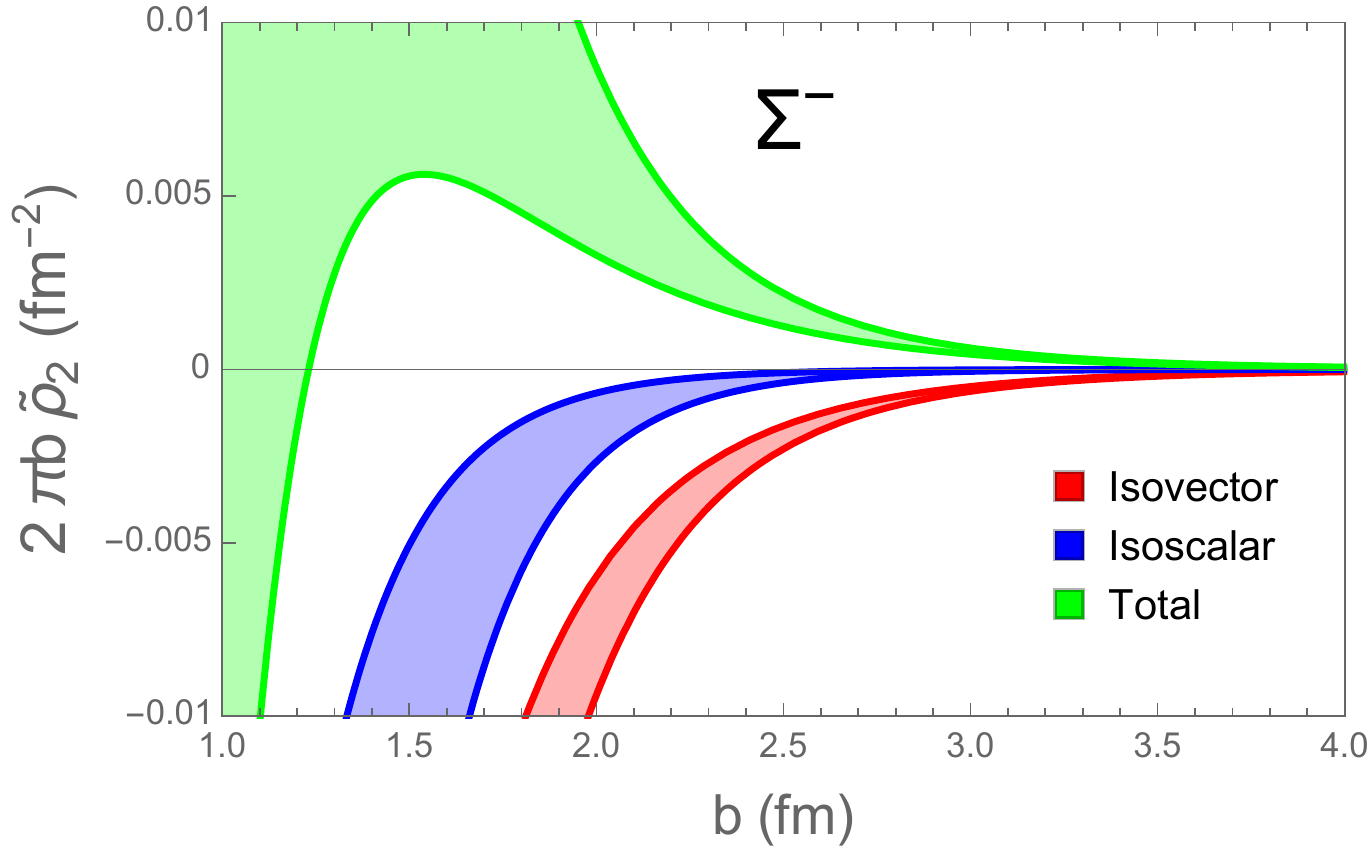,width=.45\textwidth,angle=0}\\ 
\epsfig{file=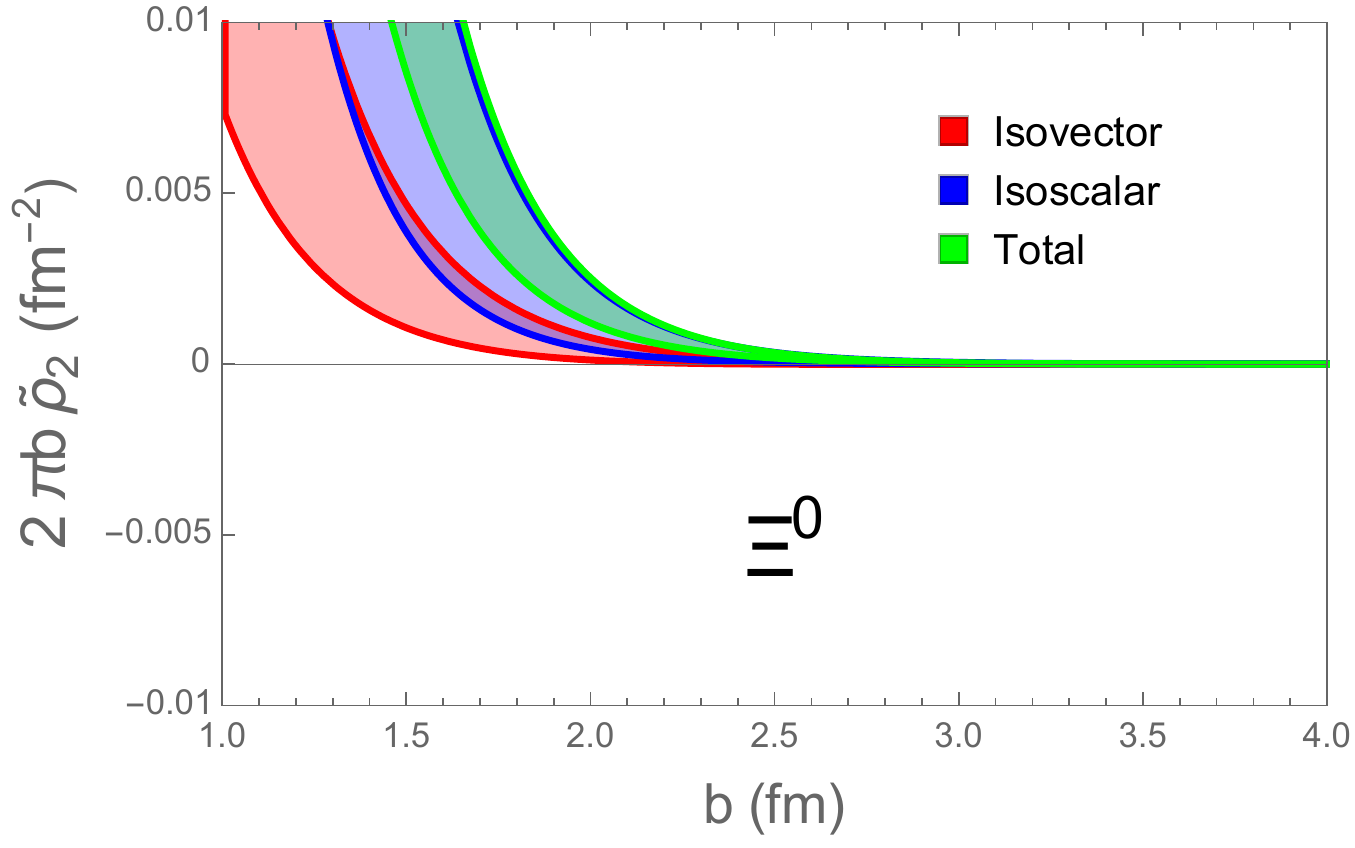,width=.45\textwidth,angle=0}
\epsfig{file=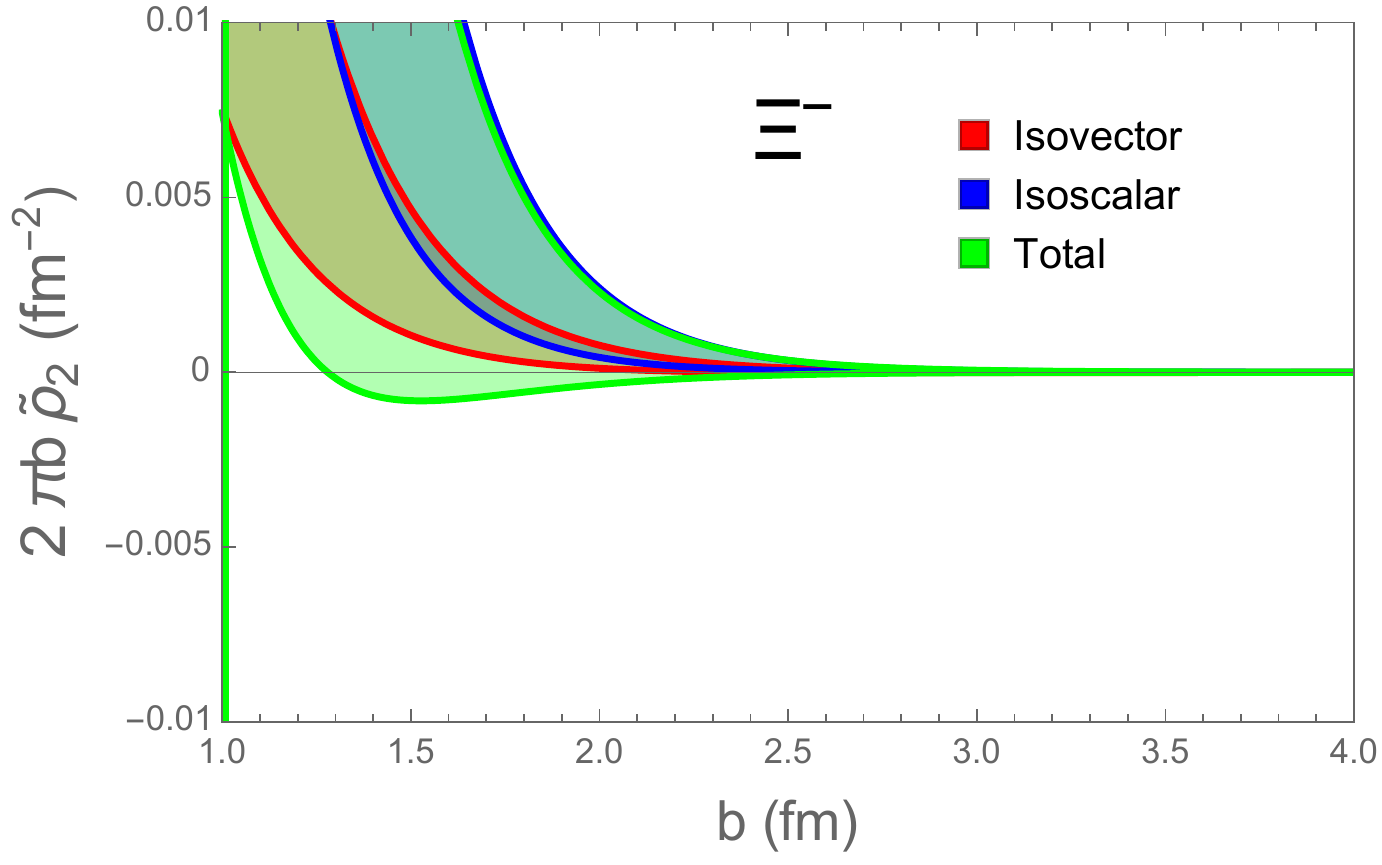,width=.45\textwidth,angle=0}\\
\epsfig{file=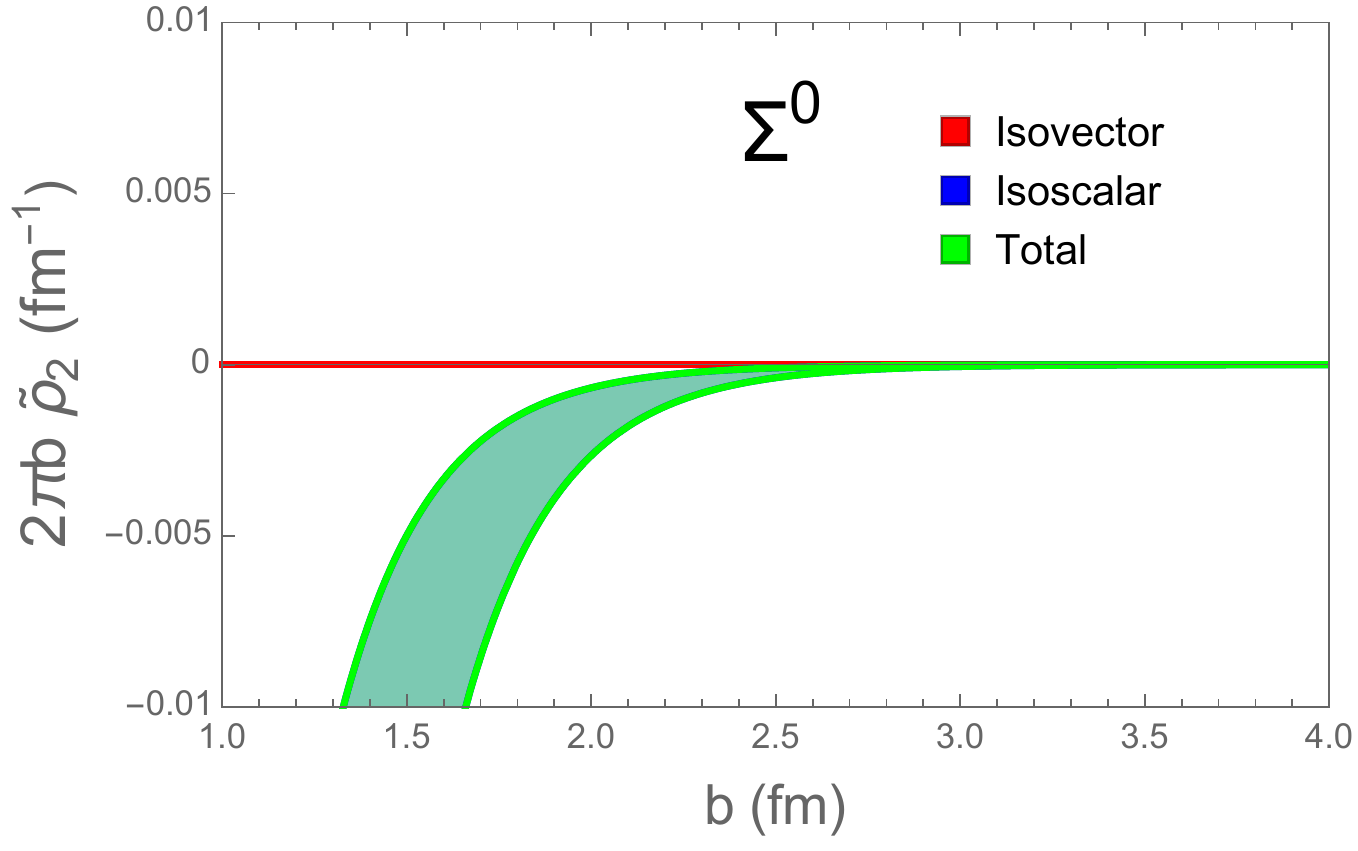,width=.45\textwidth,angle=0}
\epsfig{file=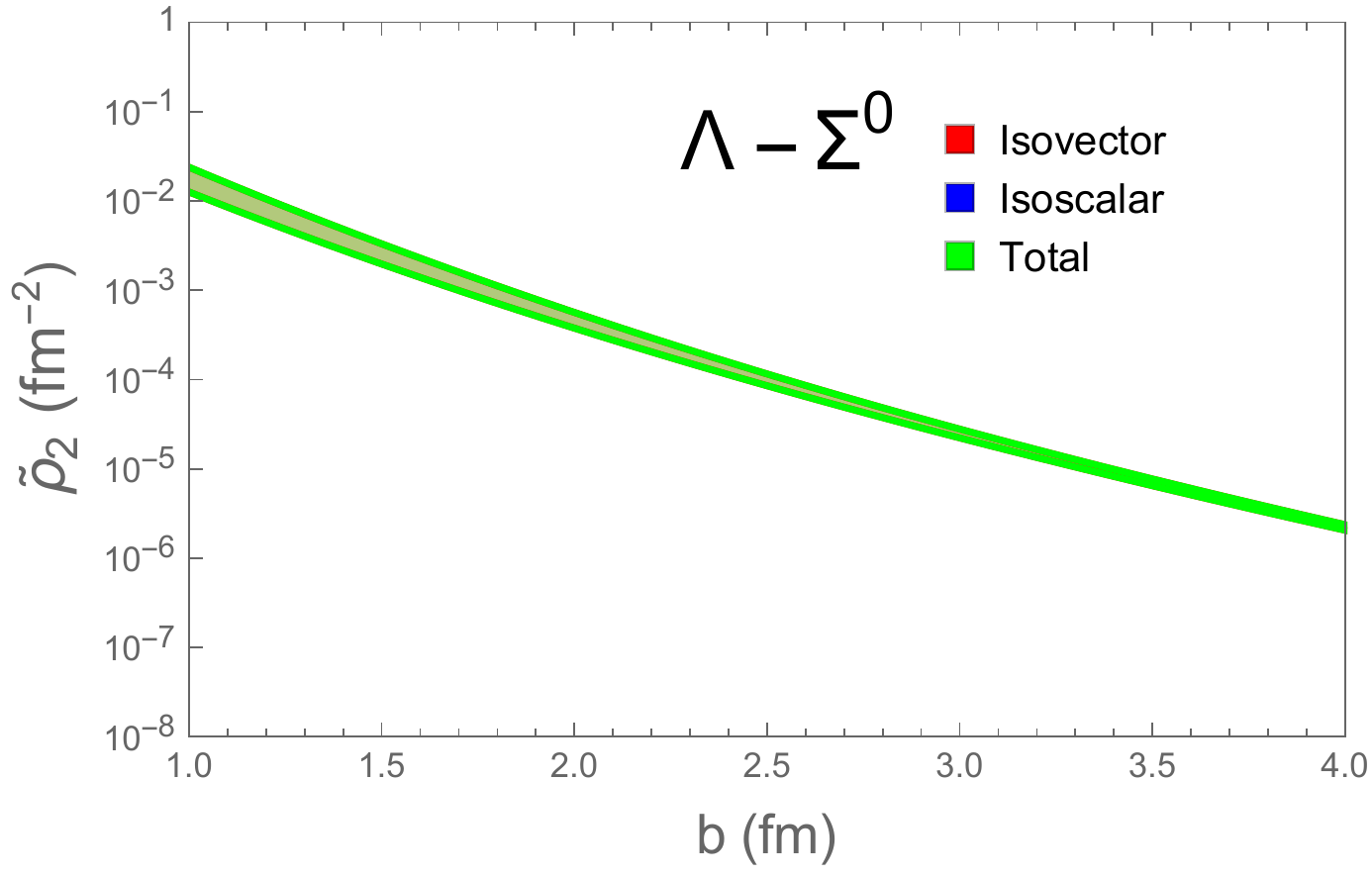,width=.45\textwidth,angle=0}
\caption[]{\small Peripheral transverse magnetization densities of the octet baryons. 
For color coding and explanations see Fig.~\ref{Fig:rho1-Octet}.}
\label{Fig:rho2-Octet}
\end{center}
\end{figure*} 
The results for the hyperon charge densities $\rho_1^B(b)$ and magnetization densities 
$\widetilde\rho_2^B(b)$
are summarized in Figs.~\ref{Fig:rho1-Octet} and \ref{Fig:rho2-Octet}. 
The baryons in the octet representation of $SU(3)$ form four isospin multiplets
\be
\left. 
\begin{array}{ccc}
\textrm{multiplet} & \textrm{baryons} & \textrm{isospin} \\
N & p, n & I = {\textstyle\frac{1}{2}} 
\\
\Lambda & \Lambda & I = 0
\\
\Sigma & \Sigma^+, \Sigma^-, \Sigma^0 & I = 1
\\
\Xi & \Xi^0, \Xi^- & I = {\textstyle\frac{1}{2}}. 
\end{array}
\hspace{1em} \right\}
\ee
Within each multiplet we write the densities as the sum/difference of an isoscalar 
and isovector component,
\be
\left.
\begin{array}{rcl}
\{ \rho^p, \rho^n \} &=& \rho^{N, S} \pm \rho^{N, V} ,
\\
\rho^{\Lambda} &=& \rho^{\Lambda, S} ,
\\
\{ \rho^{\Sigma^+}, \rho^{\Sigma^-} \} &=& \rho^{\Sigma, S} \pm \rho^{\Sigma, V} ,
\\
\rho^{\Sigma^0} &=& \rho^{\Sigma, S} ,
\\
\rho^{\Lambda - \Sigma} &=& \rho^{\Lambda-\Sigma, V} ,
\\
\{ \rho^{\Xi^0}, \rho^{\Xi^-} \} &=& \rho^{\Xi, S} \pm \rho^{\Xi, V} .
\end{array}
\hspace{1em}
\right\}
\label{isospin_rho}
\ee
The $\Lambda$ and $\Sigma^0$ densities are pure isoscalar, while the $\Lambda$-$\Sigma^0$ 
transition form factors are pure isovector. For each hyperon $B$ we show the total densities 
as well as their isovector and isoscalar components. 

The peripheral densities decay exponentially in $b$, as expected from the analytic 
properties of Eqs.~(\ref{Eq:rho1-spectral-rep}) and (\ref{Eq:rho2tilde-spectral-rep}).
The decay rate is determined by the effective $t$-values in the spectral integral. 
The isovector densities decay approximately as $\sim \exp(- M_\rho b)$ at $b \sim 1$ fm, 
and with a smaller effective mass at $b > 2 \, \textrm{fm}$, because at larger $b$ the spectral
integral shifts to smaller $t$-values closer to the two-pion threshold. The isoscalar densities
always decay with the $\omega$ mass. As a consequence, the overall densities are dominated 
by the isovector component at distances $b > 3$~fm. Since this component can be calculated
model-independently, we are able to predict the overall densities in this region within 
our approach \cite{Alarcon:2017asr}. Isoscalar and isovector densities become comparable only 
at distances $b < 2$ fm. In this region the uncertainties of our isovector calculation become 
larger, and the model dependence of the isoscalar component is significant.

The charge densities of the $\Sigma^+$ and $\Sigma^-$ show very similar behavior to
those in the proton and neutron. In both cases isovector and isoscalar components 
are present. In the $p$ and $\Sigma^+$ the isovector and isoscalar contribute
with the same sign, while in the $n$ and $\Sigma^-$ they contribute with different sign.
This gives rise to a uniform charge density in the $p$ and $\Sigma^+$, and to a more 
complex behavior in the $n$ and $\Sigma^-$. A sign change in the neutron charge density,
from negative at large $b$ to positive at $b \sim 1$ fm, was observed in the empirical 
densities \cite{Miller:2007uy,Miller:2011du}. Our results are consistent with this 
finding, but the present accuracy does not allow us to predict the sign at $b < 2$ fm. 
A similar sign change might be present in the $\Sigma^-$ density.

Our calculation shows that the peripheral isovector charge density in the charged $\Sigma$ 
multiplet is very close to that in the nucleon multiplet. This comes about due to two circumstances:
(a)~the isospin factors in the $\pi BB'$ couplings entering in the Born graphs of
Fig.~\ref{fig:eft}; (b) the relative contribution of intermediate octet and decuplet
states; see Ref.~\cite{Alarcon:2017asr} for a detailed discussion. Similar behavior is 
observed in the $\Sigma$ magnetization densities \cite{Alarcon:2017asr}. Overall this shows
that the ``pion cloud'' in the charged $\Sigma$ hyperons is of comparable size as in the nucleon.

In the $\Xi$ hyperons the peripheral isovector charge density is substantially smaller 
than in the nucleon and charged $\Sigma$ states. The reason is that the intermediate octet 
contribution to the $\Xi$ Born graphs is small and comparable to the decuplet one. 
The isoscalar density in the $\Xi$ is of normal size. This has interesting implications
for the charge density in the $\Xi^-$, which is the difference of the isoscalar and 
isovector components. It suggests a sign change from a negative charge density at large $b$
to a positive one at intermediate $b$, similar to the neutron and $\Sigma^-$, but with the 
transition occuring at larger $b$ than in the neutron or $\Sigma^-$ (we cannot confirm 
this behavior with the present uncertainties). Similar behavior is observed in 
the $\Xi$ magnetization densities \cite{Alarcon:2017asr}. Overall this means that
the ``pion cloud'' in the $\Xi$ is substantially smaller than in the nucleon and 
charged $\Sigma$.

In the $\Lambda$ and $\Sigma^0$ densities the isovector component is absent in both 
the charge and the magnetization densities, see Eq.~(\ref{isospin_rho}). In the $\chi$EFT
calculation this comes about through the cancellation of the $\pi^+$ and $\pi^-$
contributions in the Born graphs with intermediate octet and decuplet states.
The $\Lambda$ and $\Sigma^0$ densities are thus pure isoscalar, and dominated by 
$\omega$ and $\phi$ exchange in the whole range considered. This has as consequence that 
the peripheral densities are overall an order of magnitude smaller than for the 
other hyperons at $b > 2$ fm. The $\Lambda$ and $\Sigma^0$ are therefore more 
compact objects than the other hyperons as far as electromagnetic structure is concerned.
(We note that isospin symmetry breaking would resulting in a small long-range component of the
$\Lambda$ and $\Sigma^0$ densities and qualitatively change their asymptotic 
behavior \cite{Alarcon:2017asr}.)
The charge densities have the same sign for both $\Lambda$ and $\Sigma^0$,
while the magnetization densities have opposite sign.

The $\Lambda$-$\Sigma^0$ transition densities are of particular interest because of their 
pure isovector nature. They receive sizable peripheral contributions from the chiral 
processes with octet and decuplet intermediate states. These densitities can be computed
model-independently down to distances $b \sim 1$ fm and represent genuine predictions
of our approach. It would therefore be interesting to compare our results to those of other 
approaches that describe baryon structure in the central region $b < 1$ fm, such as quark 
models. The electromagnetic form factors of the hyperons are being studied also in 
Lattice QCD~\cite{Lin:2008mr,Shanahan:2014uka,Shanahan:2014cga}. If such calculations 
could determine the transverse densities in a region where both our and their approach 
are reliable, the results could be matched directly. Note also that the 
$\Lambda$-$\Sigma^0$ transition form factor is accessible experimentally through 
the Dalitz decay $\Sigma^0 \rightarrow 
\Lambda e^+e^-$ at timelike momentum transfers $4 m_e^2 < t < (m_{\Sigma^0} - m_\Lambda)^2 = 0.006 \, 
\textrm{GeV}^2$ \cite{Lutz:2009ff}. 
Such measurements may be able to determine a combination of the slopes of the 
magnetic and electric transition form factors at $t = 0$ (magnetic and electric radii),
which could be compared with dispersive calculations using the spectral functions computed 
in our approach \cite{Alarcon:2017asr} and Ref.~\cite{Granados:2017cib}.

Using similar methods one can determine also the quark flavor decomposition of
the transverse densities in the hyperon states \cite{Alarcon:2017asr}. This analysis 
requires additional assumptions about the quark composition of the isoscalar $\omega$ 
and $\phi$ exchanges (ideal mixing) and is more model-dependent. The ratios of the 
flavor densities show the transition from the ``pion cloud'' at $b >$ 3 fm
to a ``mean-field picture'' of valence quarks at $b \sim 1$ fm, as observed earlier
in the nucleon densities \cite{Miller:2011du}. These results can be used to
further quantify the ``pion cloud'' in hyperons; see Ref.~\cite{Alarcon:2017asr}
for details.
\section{Summary}
\label{Sec:Summary}
Transverse densities enable a model-independent definition of peripheral baryon
structure and its dynamical content. Using a new method combining $\chi$EFT and 
dispersion analysis, we have computed the peripheral isovector charge and 
magnetization densities in hyperons resulting from the two-pion cut of the form factors. 
The method includes $\pi\pi$ rescattering and the $\rho$ resonance and allows us to 
construct the isovector densities with controled accuracy down to $b \sim 1$ fm. 

Our results show that the ``pion cloud'' in the charged $\Sigma$ hyperons is generally
as large as that in the nucleon, while that in the $\Xi$ hyperons is substantially smaller.
The pattern is caused by the isospin factors in the $\pi BB'$ couplings and the
relative contribution of intermediate octet and decuplet states in the Born graphs.
Detailed tests of the dynamics can be performed by studying different $b$-regions
and comparing charge and magnetization densities. The $\Lambda$--$\Sigma^0$ transition 
density is pure isovector and represents a clean expression of peripheral two-pion dynamics.

The results reported here were obtained using LO $\chi$EFT.
Calculations in the $SU(2)$ sector at NLO and partial N2LO 
accuracy and show good convergence in higher orders \cite{Alarcon:2017lhg}.
This will allow us to substantially reduce the theoretical uncertainty in the 
predicted transverse densities. The method can also be extended to other baryon
form factors, such as the scalar form factor \cite{Alarcon:2017ivh}.

\section*{Acknowledgements} 

This material is based upon work supported by the U.S.~Department of Energy, 
Office of Science, Office of Nuclear Physics under contract DE-AC05-06OR23177, 
and by the Deutsche Forschungsgemeinschaft DFG. It was also funded by MINECO (Spain) and the ERDF (European Commission) grants No. FIS2014-51948-C2-2-P and SEV-2014-0398, and by the Generalitat Valenciana under Contract PROMETEOII/2014/0068.

\end{document}